# Electron-electron scattering processes in quantum wells in a quantizing magnetic field: I. Intrasubband scattering*

M.P. Telenkov, Yu.A. Mityagin

P.N. Lebedev Physical Institute of the Russian Academy of Sciences, 119991, Moscow, Russia.

Electron–electron scattering processes in a quantum well in a quantizing magnetic field are considered. A matrix of electron–electron scattering rates containing all types of transitions between Landau levels within a single subband is calculated. This matrix is analyzed, and the relative magnitude of the rates of transitions of different types is determined.

## 1. Introduction

A quantizing magnetic field alters the nature of an electron's energy spectrum in a quantum well. Continuous two-dimensional subbands become discrete sets of Landau levels, each exhibiting a high degree of degeneracy with macroscopic multiplicity [1]. This significant transformation in the energy spectrum leads to notable changes in the scattering and relaxation processes [2-15].

In a quantizing magnetic field, the Landau energy exceeds the width of levels, which significantly suppresses one-electron processes of interlevel elastic and quasi-elastic scattering (on impurities, roughness of heterointerfaces, acoustic phonons, etc.). At the same time, the intensity of electron-electron scattering increases due to the high density of states at the Landau levels. As a result, electron-electron scattering becomes the dominant mechanism for redistributing electrons across Landau levels, which determines the physical picture of relaxation processes [12,15].

Despite the principal role of electron-electron scattering, the electron-electron scattering rate matrix remains poorly explored. Only a few studies [4, 12, 15] consider a limited number of transition types resulting from electron-electron scattering.

There are, however, many more possible kinds of transitions. One of the distinctive features of Landau levels in quantum wells is their equidistant spacing. This equidistance allows for various intrasubband transitions, where the energy conservation law is satisfied regardless of the magnetic field strength. Transitions can occur when electrons are at the same Landau level (Fig. 1a) or different levels. For instance, one type of transition involves an electron from a lower level moving down while another electron from a higher level moves up the Landau level ladder (Fig. 1b). Conversely, another transition occurs when an electron with lower energy ascends while an electron with higher energy descends (Fig. 1c). Additionally, electrons may scatter to neighboring Landau levels or "jump" across multiple levels (Fig. 1d). Among these latter transitions are "crossing transitions," where the electron with lower energy in an interacting pair ends up at a higher position on the Landau level ladder after scattering than the other electron (Fig. 1e).

When Landau levels of different subbands are involved in the electron-electron scattering process, the variety of transitions increases significantly. Transitions are possible when, in the initial state, the electrons of the interacting pair are at one or different Landau levels of the same subband, and each goes to another subband (Figs. 2a and 2b).

Additionally, electron-electron scattering can take place when only one electron moves to a different subband while the other electron traverses the ladder of Landau levels within its original subband (shown in Fig. 2c). Transitions due to interactions between electrons in different subbands are also possible. In this scenario, the electrons can either pass to different subbands (Fig. 2d) or remain within their original subbands (Fig. 2e).

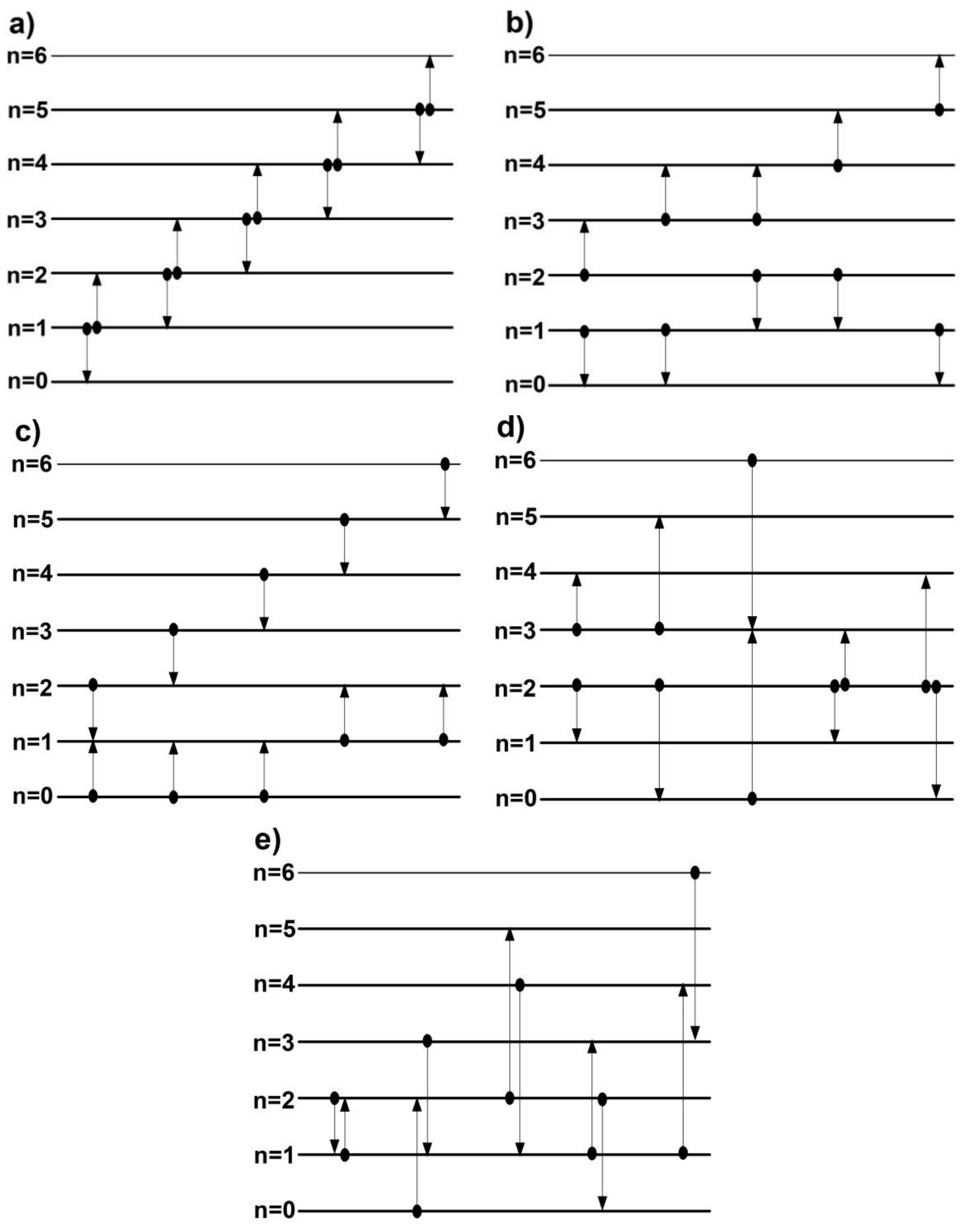

Figure. 1. Different types of interlevel transitions at electron-electron scattering in the Landau level system of one subband.

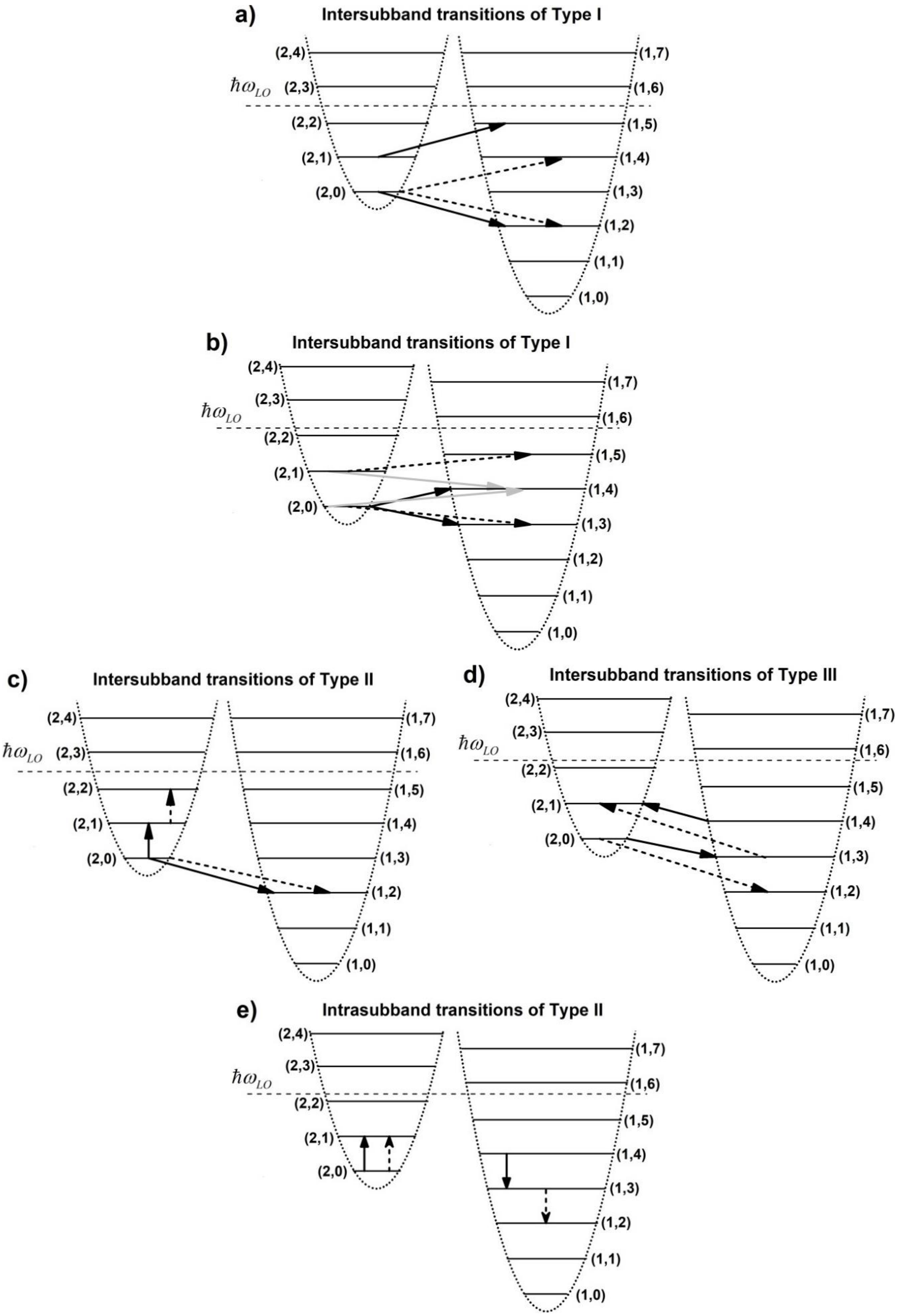

Figure 2: Different types of transitions involving Landau levels of different subbands. Arrows of different types indicate different transitions of the first and second electrons of the interacting pair.

It is important to note that the scattering rate matrix is four-dimensional. Additionally, due to the significant degeneracy factor of the Landau levels in a quantizing magnetic field, each element of this matrix represents a 10-fold integral. Our experience [11-15] in modeling relaxation processes in quantum well structures has demonstrated that hundreds of transitions of various types are necessary to describe electron kinetics accurately. As a result, calculating this matrix is a complex and time-consuming task.

This work aims to analyze the four-dimensional matrix of electron-electron scattering rates and to determine the relative magnitudes of the scattering rates for different types of transitions.

## 2. Electron spectrum in a quantum well in a quantizing magnetic field

The nature of the magnetic field effect on the electron spectrum depends essentially on its orientation relative to the axis of quantum well growth (the plane of the layers) [16].

When the magnetic field is directed parallel to the layers of the structure (perpendicular to the axis of its growth), it does not change the structure of the energy spectrum. The energy spectrum remains a set of continuous two-dimensional subbands. The influence of the magnetic field in this case is reduced to a shift of the subbands relative to each other [17] and a change in their dispersion law [18]. With an increase in the magnetic field, a flattening of the bottom of the subband occurs in the direction of the momentum component perpendicular to the magnetic field and the growth axis of the quantum well. Only at very large values of the magnetic field strength (the magnetic length is much smaller than the width of the quantum well), the electron spectrum transforms to a set of one-dimensional Landau subbands corresponding to the bulk material [19].

The situation changes substantially when the magnetic field is applied perpendicularly to the layers of the quantum well (along the axis of its growth). In this case, the magnetic field changes the character of the energy spectrum - it transforms each subband into a discrete set of Landau levels [18]:

$$E_{(\nu,n)} = \varepsilon_\nu + \hbar\omega_c\left(n+\frac{1}{2}\right), \tag{1}$$

where $\varepsilon_\nu$ - subband energies, $\omega_c = \dfrac{eB}{mc}$ - cyclotron frequency, n=0,1,2,3,… - Landau level number.

Thus, quantization arises precisely when the magnetic field is applied along the growth axis of the structure (perpendicular to its layers). The system of Landau levels in the subband is equidistant. The distance between neighboring Landau levels in the subband (Landau energy) increases proportionally to the strength of the magnetic field B and, in quantizing magnetic fields, reaches values exceeding the width of Landau levels (for example, in GaAs/AlGaAs quantum wells, the Landau energy is $\hbar\omega_c = 1.72[meV/T]\times B$ , while the characteristic width of the Landau level is ~1 meV).

Landau level degeneracy

$$\alpha = \frac{1}{\pi\ell^2} = \frac{e}{\pi\hbar c}\cdot B = 4.9\cdot 10^{10}[cm^{-2}/T]\times B \qquad (2)$$

grows linearly with the magnetic field strength and reaches macroscopic values in quantizing magnetic fields.

In the case when, in addition to the quantizing magnetic field $\mathbf{B}_\perp = B_\perp \mathbf{e}_z$ (z-axis of the structure growth), a magnetic field $\mathbf{B}_\parallel = B_\parallel \mathbf{e}_y$ parallel to the layers of the quantum well is applied (i.e., the total magnetic field $\mathbf{B} = B_\perp \mathbf{e}_z + B_\parallel \mathbf{e}_y$ is tilted to the plane of the quantum well layers), the Hamilton operator of the parabolic approximation of the envelope function formalism [21]

$$\hat{H} = \left(\hat{\mathbf{p}} + \frac{e}{c}\mathbf{A}\right)\frac{1}{2m(z)}\left(\hat{\mathbf{p}} + \frac{e}{c}\mathbf{A}\right) + U(z) \qquad (3)$$

takes in the Landau gauge $\mathbf{A} = (B_\parallel z - B_\perp y)\mathbf{e}_x$ , the form

$$\hat{H} = \hat{\mathbf{p}}\frac{1}{2m(z)}\hat{\mathbf{p}} + U(z) + \frac{m_w}{m(z)}\left[\left(\omega_\parallel z - \omega_\perp y\right)\hat{p}_x + \frac{m_w}{2}\left(\omega_\parallel z - \omega_\perp y\right)^2\right]. \qquad (4)$$

Here $U(z)$ - the potential profile of the quantum well, $m(z)$ - the effective electron mass ($m_w$ in the well and $m_b$ in the barrier), $\omega_\perp = eB_\perp / m_w c$ and $\omega_\parallel = eB_\parallel / m_w c$ - the cyclotron frequencies for the magnetic field components along the growth axis of the structure ($B_\perp$) and in the plane of its layers ($B_\parallel$) respectively.

Since $\hat{H}\hat{p}_x - \hat{p}_x\hat{H} = 0$ then we can construct a basis of stationary states with a particular momentum projection value $p_x = \hbar k_x$ on the x-axis. The wave functions of such a basis have the form

$$\Psi(x,y,z) = \frac{\exp(ik_x x)}{\sqrt{L}}\psi\left(y - k_x\ell_\perp^2, z\right), \qquad (5)$$

where $\ell_\perp = \sqrt{\hbar / m\omega_\perp} = \sqrt{\hbar c / eB_\perp}$ is the magnetic length for the transverse component of the magnetic field. The electron energy levels and wave functions of stationary states are defined by the two-dimensional Hamiltonian [22]

$$\hat{H}_{2D} = \hat{H}_\perp + \hat{H}_\parallel, \qquad (6)$$

where

$$\hat{H}_\perp = -\frac{\partial}{\partial z}\frac{\hbar^2}{2m(z)}\frac{\partial}{\partial z} + U(z) + \frac{m_w}{m(z)}\left[\frac{\hat{p}_y^2}{2m_w} + \frac{m_w\omega_\perp^2}{2}y^2\right] \qquad (7)$$

is the electron Hamiltonian when the magnetic field is directed along the growth axis of the structure. The term

$$\hat{H}_\parallel = +\frac{m_w}{m(z)}\frac{m_w\omega_\parallel^2 z^2}{2} - \frac{m_w}{m(z)}m_w\omega_\parallel\omega_\perp zy \qquad (8)$$

is due to the magnetic field component being parallel to the layers of the quantum well.

The variables in the Schrödinger equation with the Hamiltonian (7) are separable. The energy levels have the form (1), and the wave functions are given by the expression [20]

$$\Psi(x,y,z) = \frac{\exp(ik_x x)}{\sqrt{L}}\varphi_\nu(z)\Phi_n(y), \qquad (9)$$

where L is the transverse dimension of the heterostructure, $\varphi_\nu(z)$ is the wave function of the subband level $\varepsilon_\nu$ (eigenvalue wave function of the Hamiltonian $\hat{H}_z = -\frac{\partial}{\partial z}\frac{\hbar^2}{2m(z)}\frac{\partial}{\partial z} + U(z)$), $\Phi_n(y)$ - the wave function of the n-th (n=0,1,2,...) energy level of a linear harmonic oscillator with cyclotron frequency $\omega_\perp$.

Note that the ratio $m_w / m(z)$ in the third term of the Hamiltonian (7) effectively lowers the barrier height by increasing the Landau level number n and adding to $\hat{H}_z$ the term $-\left(1-\frac{m_w}{m(z)}\right)\hbar\omega_\perp\left(n+\frac{1}{2}\right)$. This leads to a dependence on n of the subband energy levels $\varepsilon$ and their wave functions $\varphi(z)$ [22]. This effect is similar in nature to dimensionality transformation at large wave vector values in quantum wells [23]. However, it is known that the dependence of the stationary states in a quantum well on the barrier height is significant only for levels close to the continuous spectrum. Therefore, we will further neglect this effect due to its smallness for the lower subbands in the considered sufficiently deep (several hundreds of meV) and wide (more than 5 nm) quantum wells at reasonable values of Landau level numbers.

The matrix of the Hamiltonian (4) in the basis of wave functions (5) is diagonal on $k_x$, and the matrix element at $k_{x1}=k_{x2}$

$$\begin{aligned}&\left\langle\frac{\exp(ik_x x)}{\sqrt{L}}\psi_1\left(y-k_x\ell_\perp^2,z\right)\middle|\hat{H}\middle|\frac{\exp(ik_x x)}{\sqrt{L}}\psi_2\left(y-k_x\ell_\perp^2,z\right)\right\rangle=\\&=\left\langle\psi_1(y,z)\middle|\hat{H}_{2D}\middle|\psi_2(y,z)\right\rangle\end{aligned}. \tag{10}$$

does not depend on $k_x$. Therefore, in the tilted quantizing magnetic field, the Landau level degeneracy is determined only by the quantizing component $B_\perp$ of the magnetic field and is given by the expression

$$\alpha=\frac{e}{\pi\hbar c}\cdot B_\perp. \tag{11}$$

The matrix element between the Landau level $(\nu_1,n_1)$ and $(\nu_2,n_2)$ is given by the expression [24]

$$\begin{aligned}&\left\langle\nu_1,n_1\middle|\hat{H}_{2D}\middle|\nu_2,n_2\right\rangle=\left[\varepsilon_{\nu_1}+\hbar\omega_\perp(n+1/2)\right]\delta_{\nu_1,\nu_2}\delta_{n_1,n_2}+\\&+\frac{m_w\omega_\parallel^2}{2}\left\langle z^2\right\rangle_{\nu_1,\nu_2}\delta_{n_1,n_2}-m_w\hbar\omega_\parallel\sqrt{\hbar\omega_\perp}\sqrt{\frac{m_w}{2\hbar^2}}\left\langle z\right\rangle_{\nu_1,\nu_2}\times\\&\times\left[\sqrt{n_2+1}\cdot\delta_{n_1,n_2+1}+\sqrt{n_2}\cdot\delta_{n_1,n_2-1}\right]\end{aligned}. \tag{12}$$

In a quantum well in a tilted magnetic field, there are two scales of energies - cyclotron energy and subband energy - the latter determined by the potential profile of the heterostructure (distance between subbands). We will be interested in the case when the cyclotron energy is several times smaller than the subband energy. In this case, in matrix (12), we can neglect the coupling between subbands (elements with $\nu_1\neq\nu_2$) and diagonalize it analytically [25-28]. As a result, the following expressions for the Landau levels and wave functions of stationary states are obtained

$$E_{(\nu,n)} = \varepsilon_\nu + \delta\varepsilon_\nu\left(B_\parallel\right) + \hbar\omega_\perp\left(n + \frac{1}{2}\right) \tag{13}$$

and

$$\Psi(x,y,z) = \frac{\exp\left(ik_x x\right)}{\sqrt{L}}\varphi_\nu(z)\Phi_n\left(y - k_x\ell_\perp^2 - \langle z\rangle_\nu tg\theta\right). \tag{14}$$

Here

$$\delta\varepsilon_\nu\left(B_\parallel\right) = \frac{m_w\omega_\parallel^2}{2}\left(\delta z\right)_\nu^2 = \frac{e^2}{2m_w c^2}\left(\delta z\right)_\nu^2 \cdot B_\parallel^2. \tag{15}$$

$$\langle z\rangle_\nu = \int dz\,\overset{*}{\varphi}(z)z\varphi(z) \tag{16}$$

is the mean value of the electron z coordinate, $\left(\delta z\right)_\nu$ - its standard deviation, $\theta$ - magnetic field tilt angle ($tg\theta = \frac{B_\parallel}{B_\perp}$).

As one can see, the Landau quantization is caused only by the magnetic field component $B_\perp$ perpendicular to the layers of the heterostructure.

The component $B_\parallel$ does not lead to the quantization of the electron energy. Its main effect on the energy spectrum is the shift of each subband by an amount proportional to $B_\parallel^2$ and $\left(\delta z\right)_\nu^2$ in the states of this subband. It also leads to the dependence of the magnetic component $\Phi\left(y\right)$ of the wave function on the subband wave function $\varphi\left(z\right)$ – there is a shift of the center of the linear harmonic oscillator by the value of $\langle z\rangle_\nu tg\theta$, proportional to $B_\parallel$.

We neglect the spin splitting of the Landau levels due to its small values in the considered structures made of III-V semiconductors of GaAs type. In the considered range of magnetic fields (1-10 T), the magnitude of the Zeeman splitting is significantly smaller than the width of the Landau level [5, 29, 30].

### 3. Electron-electron scattering rate matrix

To describe the kinetics of electrons in the Landau level system, it is necessary to calculate the electron fluxes between levels. In particular, these fluxes define a system of rate equations [7-15]

$$\frac{dN_i}{dt} = J_{e-e}^{i,+} - J_{e-e}^{i,-} + \text{single partical processes}. \tag{17}$$

Here $N_i$ is the number of electrons at the Landau level $i = (\nu_i,\ n_i)$ per unit area of the heterostructure (the population of the Landau level), $\nu_i$ is the subband number, $n_i$ is the Landau level number within the subband, $J_{e-e}^{i,-}$ is the total flux of electrons leaving Landau level i due to electron–electron scattering (the average number of outgoing electrons per unit time per unit area of the heterostructure), $J_{e-e}^{i,+}$ is the flux of electrons arriving at Landau level i as a result of electron–electron scattering.

In a rarefied electron gas (two-dimensional carrier concentration $N \ll \alpha$) in the first non-vanishing order of the RPA formalism [4]

$$J_{e-e}^{i,-} = \sum_{j,f,g} J_{e-e}\begin{pmatrix} i & j \\ f & g \end{pmatrix}, \tag{18}$$

where

$$J_{e-e}\begin{pmatrix} i & j \\ f & g \end{pmatrix} = W_{e-e}\begin{pmatrix} i & j \\ f & g \end{pmatrix} \cdot N_i N_j \left[1 - \frac{N_g}{\alpha}\right]\left[1 - \frac{N_f}{\alpha}\right], \tag{19}$$

is a total intensity of electron-electron scattering from the Landau level $i$ to the Landau level $f$, and the other from the Landau level $j$ to the Landau level $g$ (this transition will be denoted as $\{(\nu_i, n_i) \rightarrow (\nu_f, n_f) \,\&\, (\nu_j, n_j) \rightarrow (\nu_g, n_g)\}$)

$$W_{e-e}\begin{pmatrix} i & j \\ f & g \end{pmatrix} = A_{e-e}\begin{pmatrix} i & j \\ f & g \end{pmatrix} \times F_{e-e}\left(E_i + E_j - E_f - E_j\right) \tag{20}$$

is the element of the electron-electron scattering rate matrix corresponding to this transition,

$$\begin{aligned} A_{e-e}\begin{pmatrix} i & j \\ f & g \end{pmatrix} &= \frac{8\pi}{\hbar}\frac{1}{L^2}\sum_{k_i}\sum_{k_j}\sum_{k_f}\sum_{k_g} \left|V_{(i,f)(g,j)}^{e-e}\left(k_i, k_f, k_j, k_g\right)\right|^2 \cdot \frac{1}{\alpha^2} = \\ &= \frac{L^2}{\cdot 2\pi^3 \hbar \alpha^2}\int dk_i dk_j dk_f dk_g \left|V_{(i,f)(g,j)}^{e-e}\left(k_i, k_f, k_j, k_g\right)\right|^2 \end{aligned} \tag{21}$$

is the amplitude of the transition,

$$\begin{aligned} V_{(i,f)(g,j)}^{e-e}\left(k_i, k_f, k_j, k_g\right) &= \int d\mathbf{r}_1 d\mathbf{r}_2 \Psi^*_{(f,k_f)}(\mathbf{r}_1)\Psi^*_{(g,k_g)}(\mathbf{r}_2) \times \\ &\times \frac{e^2}{\varepsilon|\mathbf{r}_1 - \mathbf{r}_2|}\Psi_{(i,k_i)}(\mathbf{r}_1)\Psi_{(j,k_j)}(\mathbf{r}_2) \end{aligned}, \tag{22}$$

$\varepsilon_s$ - static dielectric constant,

Since the charge carrier gas is highly rarefied ($1/\sqrt{N} >> \ell_{\perp}$), the collisional Landau level width is small compared to the thermal energy [31], and the transition width is determined by the inhomogeneous broadening of the Landau levels due to fluctuations in the quantum well width. Note that, in order to ensure the required charge carrier concentration in the quantum well at liquid-helium temperatures, the barriers in the structures under consideration are selectively doped sufficiently far from the quantum well, as a result of which the influence of the charged impurity potential is insignificant. Since in a quantizing magnetic field the parameter of inelastic scattering on acoustic phonons $\eta = \dfrac{\hbar c_s}{\ell_{\perp}\hbar\omega_c} << 1$ , where $c_s$ is the speed of sound, then at helium temperatures in the structures under consideration the contribution of acoustic phonons to the level width is also small [32,33].

Therefore, we will consider in (19) the energies of the Landau levels $E_i(a)$ as functions of the width $a$ of the quantum well. Since the carrier gas is highly rarefied, we can assume that the electrons are distributed independently in the plane of the layers. Then, averaging over the well width fluctuations leads to replacing the Dirac delta function in (20) with the form factor.

$$F_{e-e}\begin{pmatrix} i & j \\ f & g \end{pmatrix} = \int da_1 \int da_2 \int da_3 \int da_4 \Pi(a_1)\Pi(a_2)\Pi(a_3)\Pi(a_4) \times$$
$$\times \delta\left(E_i(a_1) + E_j(a_2) - E_f(a_3) - E_g(a_4)\right) \tag{23}$$

Using a normal distribution for the quantum well width in the plane of the layers,

$$\Pi(a) = \frac{1}{\sqrt{2\pi}\sigma}\exp\left(-\frac{(a-a_0)^2}{2\sigma^2}\right), \tag{24}$$

in the vicinity of the average value $a_0$ of the well width, we obtain

$$F_{e-e}\begin{pmatrix} i & j \\ f & g \end{pmatrix} = \int dE_1 dE_2 dE_3 dE_4 \rho_{i,k_i}(E_1)\rho_{f,k_f}(E_2) \times$$
$$\times \rho_{j,k_h}(E_3)\rho_{g,k_g}(E_4)\cdot\delta(E_1 + E_3 - E_2 - E_4) \tag{25}$$

where

$$\rho_{i,k_i}(E) = \int da \Pi(a)\delta\left(E - E_i(a)\right) = \frac{1}{\sqrt{2\pi}\Gamma}\exp\left(-\frac{\left(E - E_i(a_0)\right)^2}{2\Gamma^2}\right) \tag{26}$$

is the density of states of the inhomogeneously broadened Landau level, and $\Gamma = \gamma\sigma$ is the magnitude of the inhomogeneous broadening of the Landau level.

The form factor (26) is Gaussian

$$F_{e-e}\begin{pmatrix} i & j \\ f & g \end{pmatrix} = \frac{1}{\sqrt{2\pi}\Gamma_{\Sigma}^{ee}} \exp\left\{ -\frac{\left(E_i\left(a_0\right)+E_j\left(a_0\right)-E_f\left(a_0\right)-E_j\left(a_0\right)\right)^2}{2\left(\Gamma_{\Sigma}^{ee}\right)^2} \right\} \equiv \equiv F_{ee}\left(E_i+E_j-E_f-E_j\right) \tag{27}$$

with a maximum corresponding to the energy conservation law in electron-electron scattering

$$E_i\left(a_0\right)+E_j\left(a_0\right)=E_g\left(a_0\right)+E_f\left(a_0\right)$$

with an average well width $a_0$ and the transition width $\Gamma_{\Sigma}^{ee}=\sqrt{\Gamma_i^2+\Gamma_j^2+\Gamma_f^2+\Gamma_g^2}$ .

In the future, we will understand the value of the Landau level energy at the average well width $a_0$. When performing averaging, we neglected the dependence of the subband wave function $\varphi_{\nu}(z)$ on the quantum well width due to the smallness of the effect.

The scattering rate amplitude is represented as a multiple integral containing a Coulomb singularity. This complexity makes calculating it quite challenging and time-consuming, which likely explains the lack of detailed information on the rate matrix in the literature.

Using the wave functions (14), we could perform a partial integration in (21) and thereby substantially lower the multiplicity of the integral. As a result, we obtained the following expression for the electron-electron scattering amplitude:

$$A_{e-e}\begin{pmatrix} i & j \\ f & g \end{pmatrix} = \frac{4e^4}{\pi^2\varepsilon_s^2\hbar} \frac{\exp\left(\left[-\frac{(\xi_{\nu_f,\nu_i}+\xi_{\nu_g,\nu_j})^2}{4}\right]\right)}{2^{n_i+n_j+n_g+n_f}\, n_i!n_j!n_g!n_f!} \times \times\iint dk_1 dk_2 \exp\left(-\left[k_2-\frac{(\xi_{\nu_i,\nu_j}+\xi_{\nu_g,\nu_f})}{2}\right]^2\right)\left|M_{i,j,g,f}\left(k_1,k_2\right)\right|^2 , \tag{28}$$

where

$$M_{i,j,g,f}\left(k_1,k_2\right)=\int dy \exp\left(-\frac{1}{2}\left(y+k_1-k_2\right)^2\right)\times$$
$$\times G_{\nu_i,\nu_j,\nu_g,\nu_f}\left(\left|k_2\right|;\left|y+\frac{\xi_{\nu_i,\nu_j}+\xi_{\nu_f,\nu_g}}{2}\right|\right)\Lambda_{n_i,n_j,n_g,n_f}\left(k_2,y+k_1-k_2\right) \tag{29}$$

$$G_{\nu_i,\nu_j,\nu_g,\nu_f}(\gamma;y)=\int dz_1 K_0\left(\gamma\sqrt{y^2+\frac{4z_1^2}{\ell_\perp^2}}\right)\times R_{\nu_i,\nu_j,\nu_g,\nu_f}(z_1), \tag{30}$$

$$R_{\nu_i,\nu_j,\nu_g,\nu_f}(z_1)=\int dz_2 \varphi_{\nu_i}(z_2)\varphi_{\nu_j}(z_2-2z_1)\overset{*}{\varphi}_{\nu_g}(z_2-2z_1)\overset{*}{\varphi}_{\nu_f}(z_2), \tag{31}$$

$$\Lambda_{n_i,n_j,n_g,n_f}(k,y)=2^{\frac{n_i+n_j+n_g+n_f}{2}}\sqrt{2}\sum_{p_1=0}^{n_i}\frac{1}{2^{p_1/2}}\binom{n_i}{p_1}H_{p_1}(\delta_i)\sum_{p_2=0}^{n_j}\frac{1}{2^{p_2/2}}\binom{n_j}{p_2}H_{p_2}(\delta_j)\times$$
$$\times\sum_{p_3=0}^{n_g}\frac{1}{2^{p_3/2}}\binom{n_g}{p_3}H_{p_3}(\delta_g)\times\sum_{p_4=0}^{n_f}\frac{1}{2^{p_4/2}}\binom{n_f}{p_4}H_{p_4}(\delta_f)\times$$
$$\times\frac{\left[1+\left(-1\right)^{n_i+n_j+n_g+n_f-p_1-p_2-p_3-p_4}\right]}{2}\times$$
$$\times\Gamma\left(\frac{n_i+n_j+n_g+n_f-p_1-p_2-p_3-p_4}{2}+\frac{1}{2}\right) \tag{32}$$

$K_0(x)$ - McDonald function, $\Gamma(x)$ - Euler gamma function, $H_n(x)$ - Hermite polynomial of degree n,

$$\delta_i=\frac{y+k+\xi_{\nu_f,\nu_i}}{2}, \tag{33}$$

$$\delta_j=\frac{-y-k+\xi_{\nu_g,\nu_j}}{2}, \tag{34}$$

$$\delta_g=\frac{-y+k-\xi_{\nu_g,\nu_j}}{2}, \tag{35}$$

$$\delta_f=\frac{y-k-\xi_{\nu_f,\nu_i}}{2}. \tag{36}$$

$$\xi_{\nu_1,\nu_2}=\sqrt{\frac{e}{\hbar c B_\perp}}\cdot\left[\left\langle z\right\rangle_{\nu_1}-\left\langle z\right\rangle_{\nu_2}\right]\cdot B_\parallel, \tag{37}$$

It is worth noting that, in addition to significantly reducing the complexity of the integral, the resulting expression offers several advantages for numerical calculations. Firstly, all relevant information about the heterostructure is encapsulated in a single integral, specifically integral (28), which is dependent on the subband wave functions $\varphi(z)$. Moreover, this integral is defined only by $\varphi(z)$ and does not include the components of the wave function (14) along other coordinate axes. Therefore, the function $R_{\nu_i,\nu_j,\nu_g,\nu_f}(z_1)$ defined by (28) is the same for all transitions between fixed subbands $\nu_i,\nu_j,\nu_g$ and $\nu_f$ in a given heterostructure. Accordingly, it can be calculated only once and used for all transitions with possible combinations of $n_i,n_j,n_g$ and $n_f$ at any magnetic field magnitude. This reduces the multiplicity of integration by one. Furthermore, for many commonly used heterostructures, especially rectangular quantum wells, analytical expressions for the subband wave functions $\varphi(z)$ are available. This allows for the integral (28) to be calculated analytically.

The resulting expressions allow us to describe electron–electron scattering processes both within a single subband and between Landau levels of different subbands, even under a quantizing magnetic field directed at any angle relative to the structure layers.

In this paper, we present the results of a study on the scattering rate matrix elements W related to various types of intrasubband electron–electron scattering processes. As discussed in Section 2, the key features of the electron spectrum in the system we are studying, which primarily differentiate it from other electron systems, are governed by the magnetic field component aligned with the growth axis of the heterostructure. To emphasize the significance of these features, this paper examines scattering processes in a magnetic field oriented perpendicular to the heterostructure layers.

## 4. Intrasubband transitions

This section presents the results of a study on intra-subband electron–electron scattering processes, where all transitions between Landau levels occur within a single subband. The obtained regularities are illustrated using the example of the GaAs/$Al_{0.3}Ga_{0.7}$As quantum well with a width of 25 nm, unless otherwise specified. For such quantum wells, the typical inhomogeneous level broadening $\Gamma$ is approximately 1 meV. Therefore, in further calculations, we will assume $\Gamma$ = 1 meV.

Electron-electron scattering is an elastic two-particle process. For the transition $\{(\nu_i,n_i)\to(\nu_f,n_f);(\nu_j,n_j)\to(\nu_g,n_g)\}$, the energy conservation law (the transition resonance condition) has the form

$$E_{(\nu_i,n_i)} + E_{(\nu_j,n_j)} = E_{(\nu_f,n_f)} + E_{(\nu_g,n_g)} . \quad (38)$$

Substitution in (35) of the expression (13) for the single-particle energy gives

$$n_f - n_i + n_g - n_j = \frac{\left(\varepsilon_{\nu_i} - \varepsilon_{\nu_f} + \varepsilon_{\nu_j} - \varepsilon_{\nu_g}\right)}{\hbar\omega_c} . \quad (39)$$

For intrasubband transitions $\nu_i = \nu_j = \nu_g = \nu_f$ and, respectively, the resonance condition (37) takes the form

$$n_f - n_i + n_g - n_j = 0 . \quad (40)$$

The equidistance of the one-electron spectrum within a subband means that the resonance condition (the energy conservation law) for intrasubband processes is independent of the magnetic field. Consequently, the rates of intrasubband electron-electron scattering decrease gradually and slowly as the magnetic field strength increases.

The above is illustrated by Figure 3, which shows a typical dependence of the intrasubband scattering rate on the magnetic field. The slight decrease of the scattering rate can be explained by the fact that with increasing magnetic field, the magnetic length, which defines the localization area of the wave function, decreases (the RMS deviation of the coordinate at the n-th level of the linear harmonic oscillator is equal to $\ell\sqrt{n+1/2}$, $\ell = \sqrt{\hbar c / eB}$ is the magnetic length). Consequently, the overlapping of the interacting electron wave function in the plane of layers goes low, and the scattering probability decreases. Since the magnetic length changes relatively slowly with the magnetic field (~ $\sqrt{B}$ ), lowering the electron-electron scattering rate with increasing magnetic field is a relatively slow process.

As a result, at any magnetic field value, numerous transitions within the subband are possible, and for each of them the energy conservation law is satisfied. Three numbers can describe this rich set of transitions, and the following general formula for intrasubband transitions can be written as follows:

$$\left\{ n \to n - \Delta n \,\&\, n + \Delta N \to n + \Delta N + \Delta n \right\} , \quad (41)$$

The subband number, which remains constant across all levels, is omitted. Here, $n$ - is the number of the Landau level from which the electron of the scattering pair passes down the ladder of Landau levels, and $n+\Delta N$ - the number of the Landau level from which the second electron passes up the ladder of Landau levels, $\Delta n$=1,2,… is the absolute value of the change of the Landau level number of an individual electron of the scattering pair. As can be seen from (41), the transition of an electron down the ladder of Landau levels with a decrease in the level number is always accompanied by the transition of another electron up the ladder of Landau levels with an increase of the number by the same value. In this case, the first electron diminishes its energy by the value of $\hbar\omega_c\Delta n$, while the second electron enhances its energy by the same value. In the following, we will use the concept of transferred energy $E_{\text{trans}}$ in the electron-electron scattering event. We will define it as the value by which the energy of one electron increases (the energy of the other electron decreases by the same value). In the case of intrasubband scattering $E_{trans} = \hbar\omega_c\Delta n$, $\Delta n$ is the transferred energy expressed in Landau energy units.

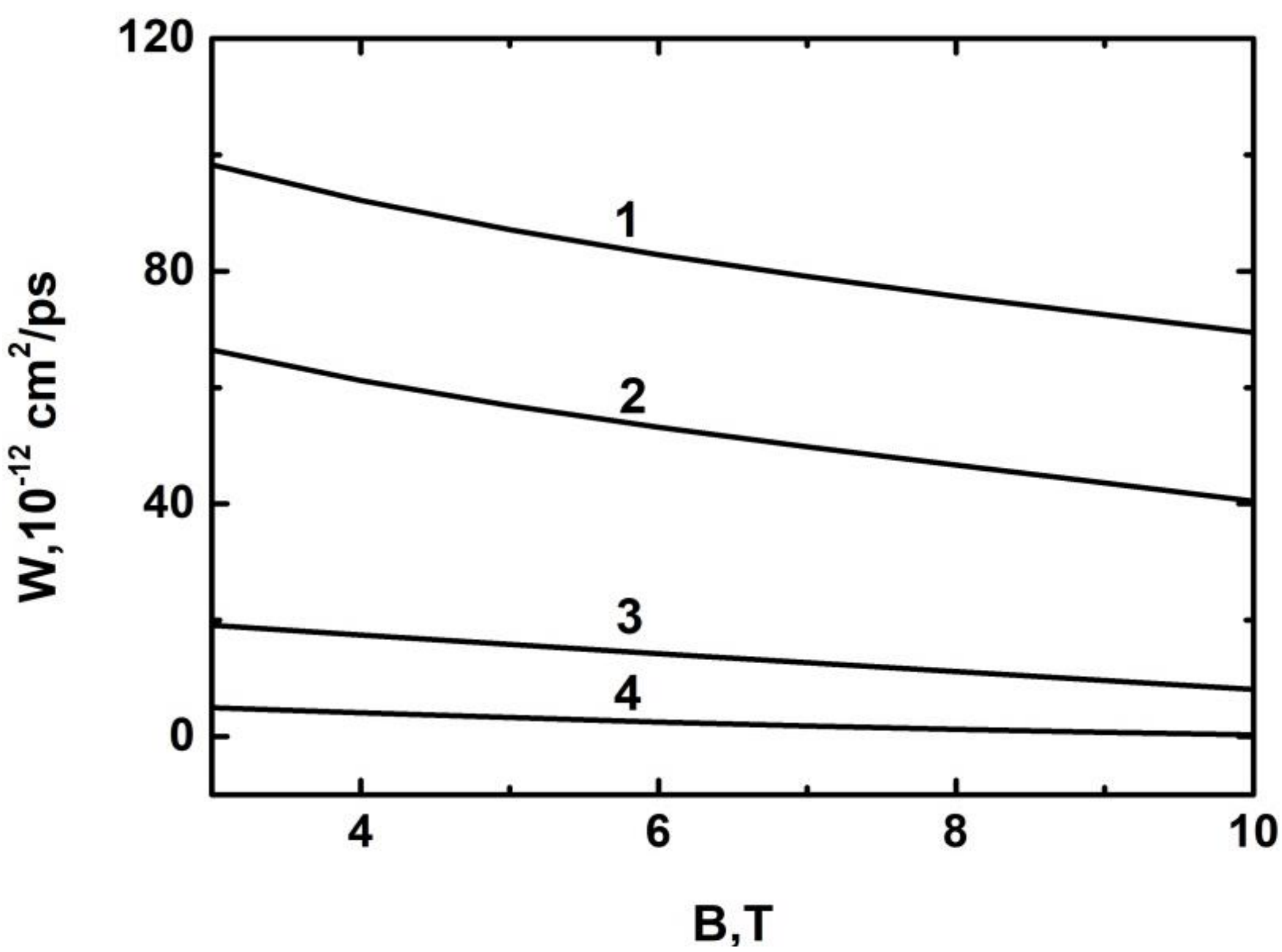

Figure 3: Dependence of the electron-electron scattering matrix elements $W\begin{pmatrix} i & j \\ f & g \end{pmatrix}$ on the magnetic field strength. 1 – i = j = (1, 3), f = (1, 2), g = (1, 4);
2 – i = (1, 1), j = (1, 2), f = (1, 0), g = (1, 3); 3 – i = j = (1, 3), f = (1, 1), g = (1, 5);
4 – i = j = (1, 3), f = (1, 0), g = (1, 6)

The value $\Delta N$ is the difference between the initial Landau level numbers of two electrons before scattering – $\Delta N=n_i-n_j$, $n_i$ is the Landau level number of the electron, which passes up the ladder of Landau levels as a result of scattering, and $n_j$ - the number of the initial Landau level of the second electron in the interacting pair. Accordingly, the difference in the energy of the first and second electrons before scattering is equal to $\Delta E_{init} = \left|E_i - E_j\right| = \hbar\omega_c \left|\Delta N\right|$. In other words, $\left|\Delta N\right|$ is a distance in the energy space between the electrons of the interacting pair in their initial state (the distance on the ladder of Landau levels).

If $\Delta N$=0, the formula (41) takes the form

$$\left\{ n \to n - \Delta n \,\&\, n \to n + \Delta n \right\}. \quad (42)$$

F

and describes transitions when, before scattering, both electrons are at the same Landau level $n$. Examples of transitions with $\Delta N$=0 are shown graphically in Figure 4a.

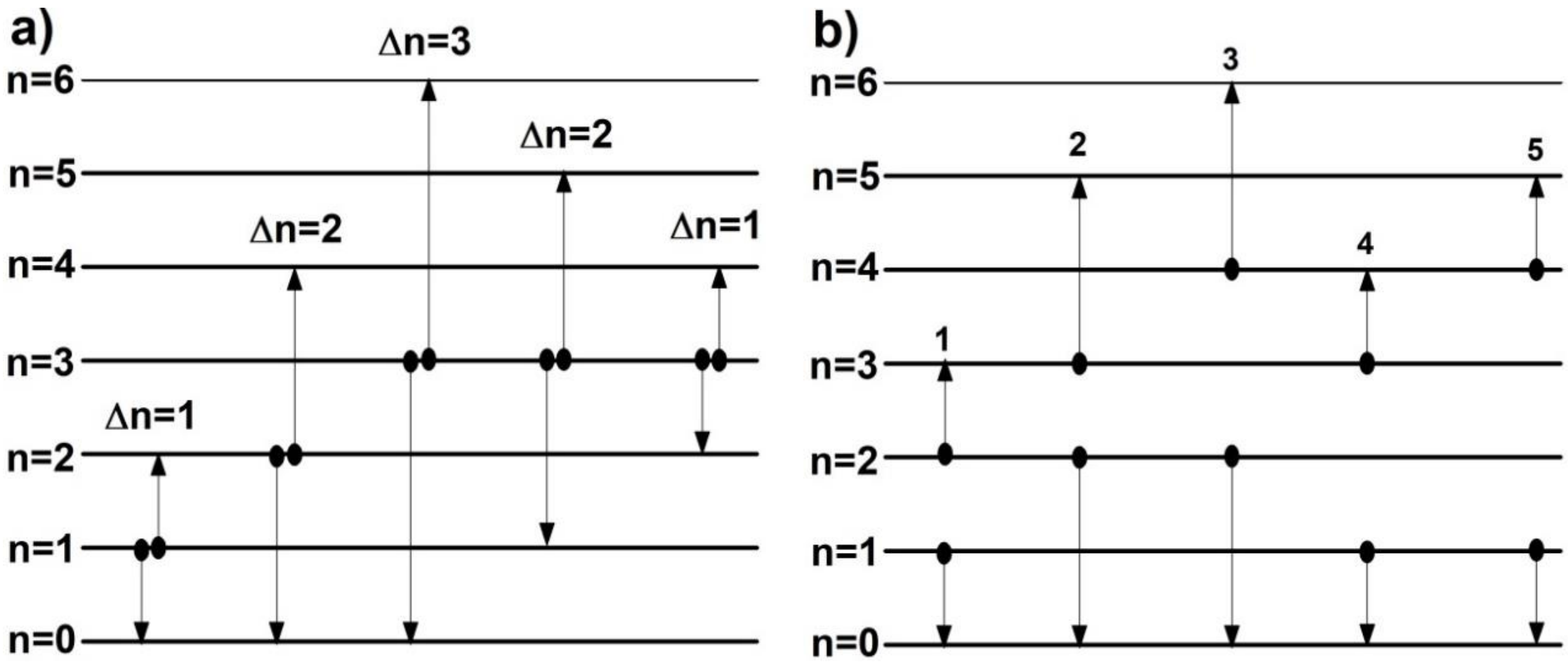

Figure. 4. Schematic of intrasubband transitions (a) with $\Delta N = 0$; (b) with $\Delta N > 0$: 1 - ($\Delta N = 1, \Delta n = 1$), 2 - ($\Delta N = 1, \Delta n = 2$), 3 - ($\Delta N = 2, \Delta n = 2$), 4 - ($\Delta N = 2, \Delta n = 1$), 5 - ($\Delta N = 3, \Delta n = 1$)

At $\Delta N$<0, on the contrary, the electron of the scattering pair passing to the higher level is initially lower on the ladder of Landau levels than the electron passing down the ladder. Thus, in this case, the electrons transit toward each other along the ladder of Landau levels. Among such transitions, we can distinguish transitions with $\left|\Delta N\right| \geq 2\Delta n$ (Fig. 5a) and transitions with $\left|\Delta N\right| < 2\Delta n$ (Fig. 5b). In the first case, the electron in the interacting pair, which has a higher energy in the initial state, also has a higher energy after scattering. In the second case, this electron has a lower energy after scattering than the second electron of the pair (crossing transitions take place).

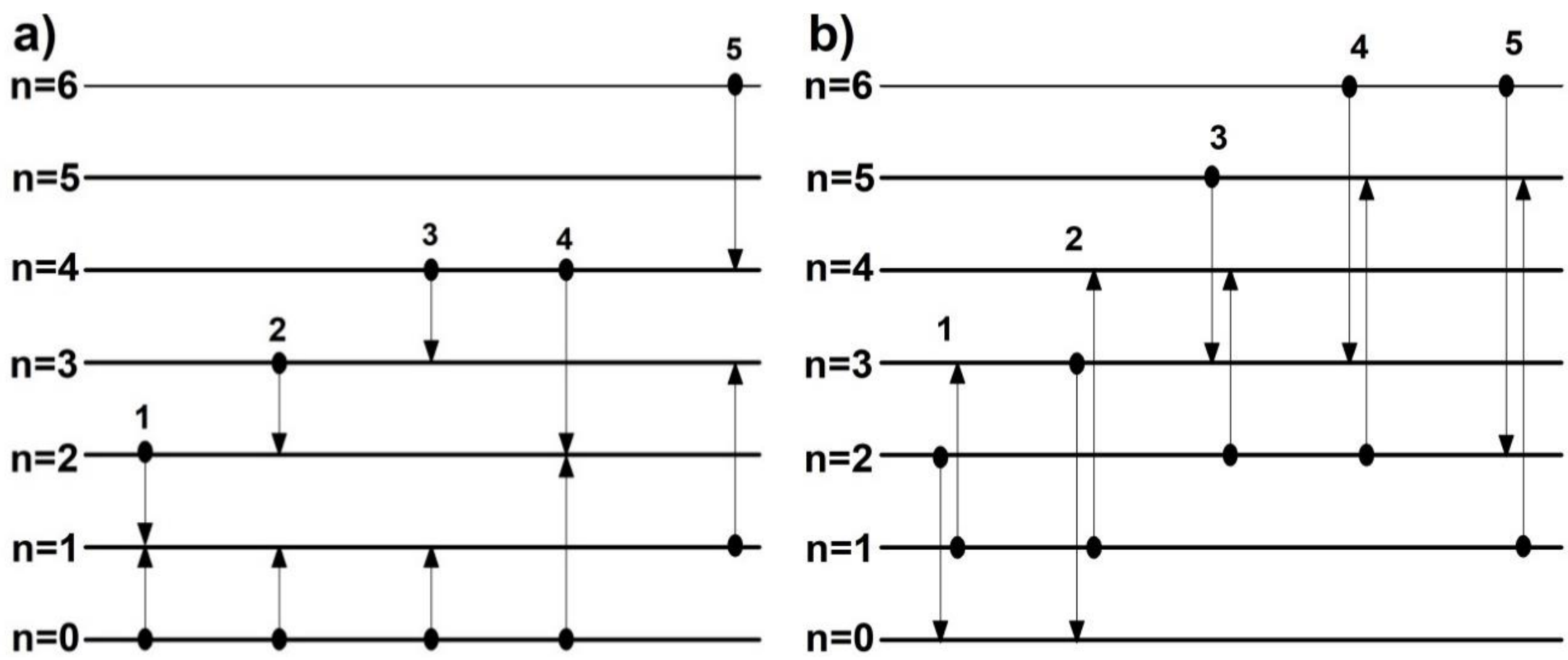

**Figure. 5.** Scheme of intrasubband transitions with $\Delta N < 0$:
a) $|\Delta N| \geq 2\Delta n$: 1 – ($\Delta N = -2, \Delta n = 1$), 2 – ($\Delta N = -3, \Delta n = 1$), 3 – ($\Delta N = -4, \Delta n = 1$), 4 – ($\Delta N = -4, \Delta n = 2$), 5 – ($\Delta N = -5, \Delta n = 2$);
б) $|\Delta N| < 2\Delta n$: 1 – ($\Delta N = -1, \Delta n = 2$), 2 – ($\Delta N = -2, \Delta n = 3$), 3 – ($\Delta N = -3, \Delta n = 2$), 4 – ($\Delta N = -4, \Delta n = 3$), 5 – ($\Delta N = -5, \Delta n = 4$)

In this work, we have considered various combinations of n, $\Delta n$ and $\Delta N$ for quantum wells with different widths, thus studying all possible types of intra-subband transitions. As a result, the following trends of the behaviour of the elements of the electron-electron scattering rate matrix have been established.

First, the rates of intra-subband transitions decrease rapidly with increasing $\Delta n$ and, consequently, with the energy $E_{trans}$ transferred to one electron. The above is illustrated by Figure 6, which shows a typical dependence of the transition rate on $\Delta n$ and, respectively, on the $E_{trans}$.

The tendency of decreasing matrix elements between stationary states as the energy difference between them increases is quite common in quantum systems. It is because as a state's energy increases, the number of zeros of the wave function increases, which leads to a decrease in the integral defining the matrix element. In particular, this circumstance is the basis of truncated basis methods, in which not the full exact basis with an infinite number of states is used, but only a finite number of states corresponding to levels falling in a finite energy range. In our case, the one-electron state $(\nu, n)$ wave function is the product of the wave function of the $\nu$ -th subband by the wave function of the nth energy level of the linear harmonic oscillator, which has n zeros. At the intra-subband transition from level n to level $n \pm \Delta n$, the subband wave function does not change, but the number of zeros in the oscillator wave function changes by $\Delta n$, which leads to a change by $\Delta n$ the number of zeros on the coordinate part of the first electron wavefunction in the matrix element integral.

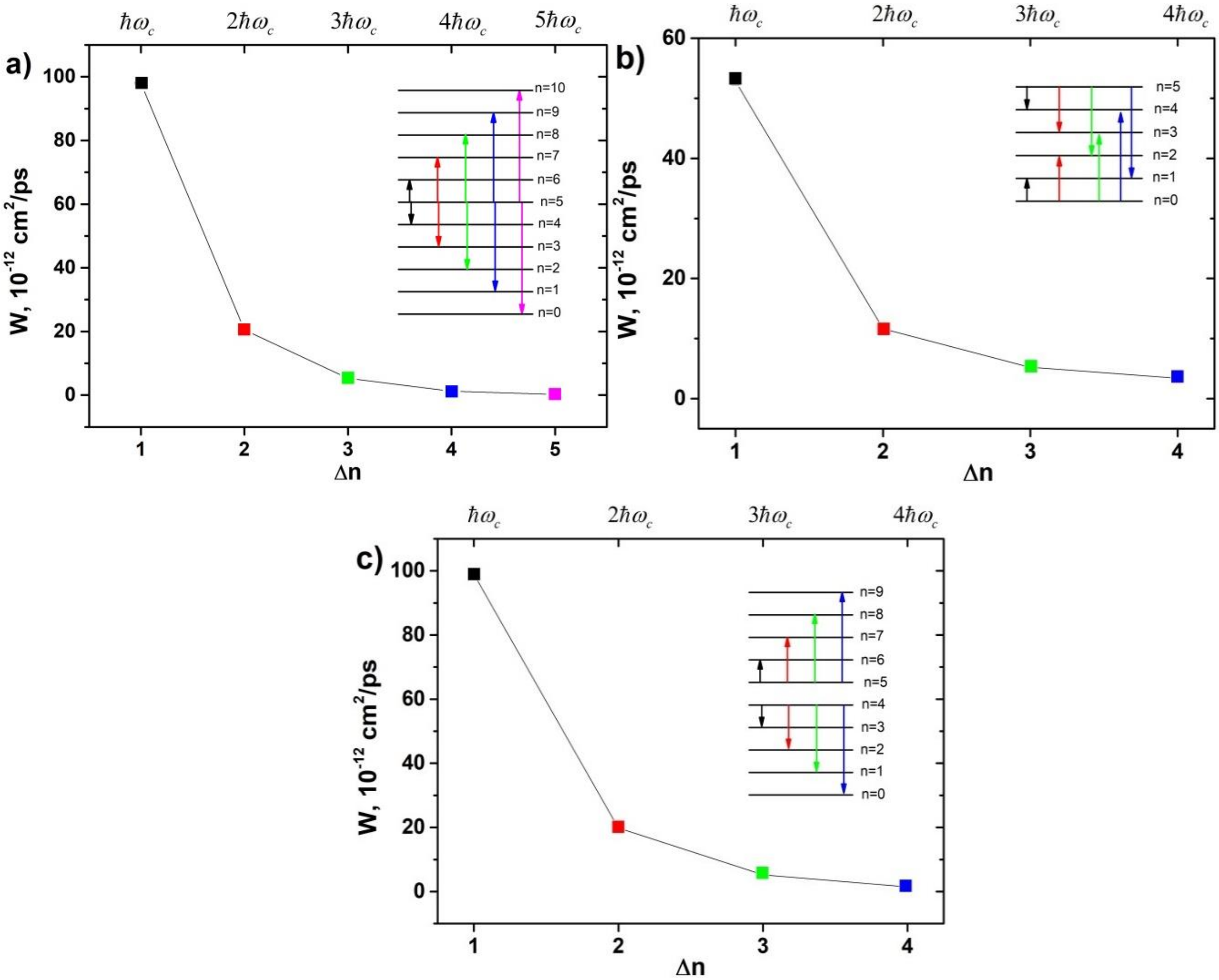

Figure. 6. Dependence of the intrasubband transition rate on the change of the Landau level number of the electron at the transition Δn (bottom axis) and, accordingly, on the transferred energy. Magnetic field strength B=3.5 T. a) $n=5, \Delta N=0$; b) $n=5$, $\Delta N=-5$; c) $n=4$, $\Delta N=1$. The inserts show the schemes of transitions. The color of the arrows indicating the transition coincides with the color of the point on the graph.

Similarly, the number of zeros in the coordinate part of the second electron wavefunction of its final state also changes by $\Delta n$. As a result, the matrix element of the transition $\{n \rightarrow n-\Delta n \,\&\, n+\Delta N \rightarrow n+\Delta N+\Delta n\}$ decreases with increasing $\Delta n$, and, accordingly, the transition rate decreases. Moreover, the difference between the transition rates with $\Delta n$ and $\Delta n+1$ decreases with increasing $\Delta n$. The explanation is that the more zeros the wave function has, the weaker it changes when one zero is added.

The dependence of the electron-electron scattering rate on the difference in the Landau level numbers of the interacting electrons and, accordingly, the difference in their energy before scattering (the difference in the energies of the interacting electrons in the initial state of the corresponding transitions) was found to be somewhat unexpected, at first glance.

With increasing $|\Delta N|$ (the difference in the energy of the electrons before scattering), the electron-electron scattering rate decreases. However, this dependence is relatively weak, so the rates of transitions in the initial states of which the electrons are at levels rather far apart on the Landau level ladder are close to the rates of transitions in which the interacting electrons are initially at the same Landau level.

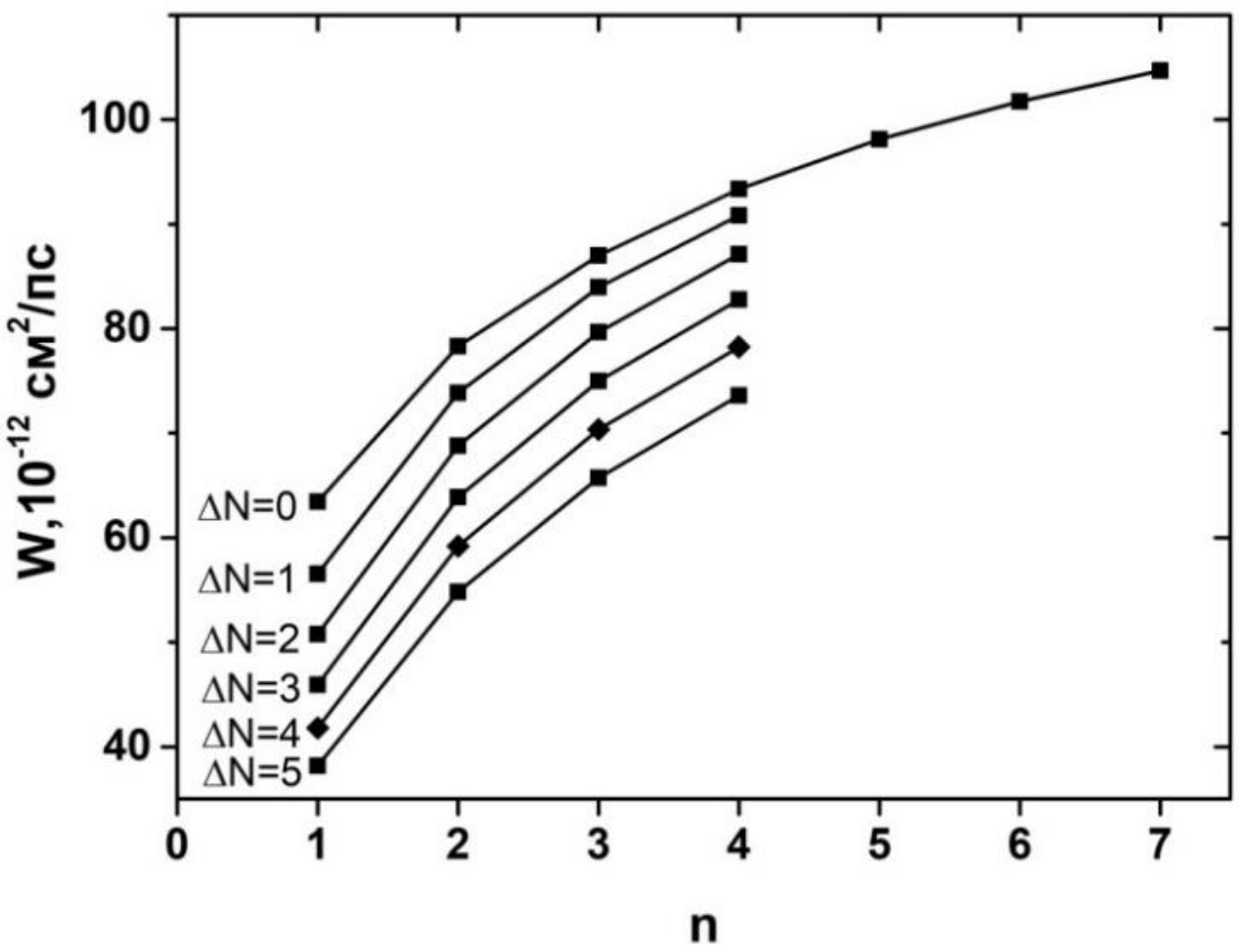

Figure 7. Dependence of the intrasubband electron-electron scattering rate on $n$ at fixed values of $\Delta n$ and $\Delta N$. Different curves correspond to different values of $\Delta N$. For all transitions $\Delta n = 1$. The data are given for the lower subband of the GaAs/$Al_{0.3}Ga_{0.7}As$ quantum well of 25 nm width. Magnetic field $B$=3.5 T.

As an illustration, Figure 8 shows a typical intrasubband electron-electron scattering rate dependence on $\Delta N$. The different curves correspond to transitions $\{n \to n-1 \,\&\, n+\Delta N \to n+\Delta N+1\}$ with varying values of *n*. For all transitions shown in Figure 8 $\Delta n = 1$. The scattering rate weakly decreases with increasing $\Delta N$.

One can explain this behavior in the following way. The difference between the wave functions of the initial states of the electrons of the scattering pair increases with increasing $\Delta N$, which leads to a decrease in the transition matrix element. However, in the matrix element integral (21), the variables $\mathbf{r}_1$ at the wave functions of the initial and final states of the first electron differ from the variables $\mathbf{r}_2$ at the wave functions of the second electron of the scattering pair (the variables at the wave functions of the first and second electron are independent). Therefore, the difference in the wave functions of different electrons (i.e., $\Delta N$) has a much weaker effect on the scattering rate than that of a single electron (i.e., $\Delta n$).

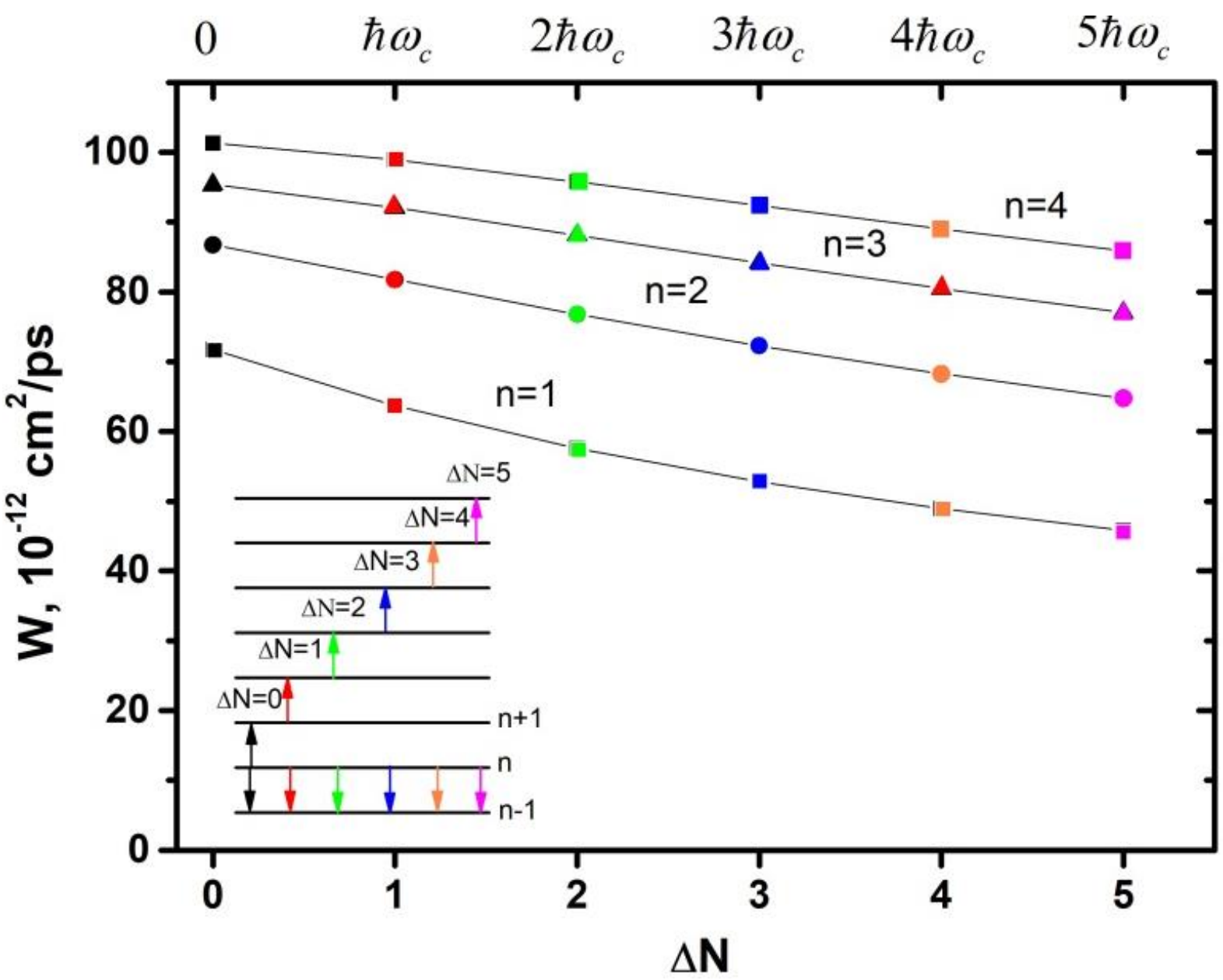

Fig. 8. Dependence of the intrasubband electron-electron scattering rate on the distance between the Landau levels from which the transition occurs (lower axis). The corresponding energy difference of the electrons of the interacting pair in the initial transition state is plotted on the upper axis. Magnetic field B=3.5 T.

In the curves presented in Figure 8, the initial and final Landau levels of one electron are fixed, while for the second electron, the number of the level from which it moves up the ladder of Landau levels is increasing. Following our "concept of zeros", this should increase the transition rate and partially compensate for the decrease in the matrix element caused by the increase in $\Delta N$.

Therefore, one would expect that in a transition sequence where the initial and final Landau levels of one electron are fixed and the second electron moves down the ladder of Landau levels, the decrease in the transition matrix element Therefore, one would expect that in a transition sequence where the initial and final Landau levels of one electron are fixed and the second electron moves down the ladder of Landau levels, the decrease in the transition matrix element with $\Delta N$ could be pretty significant

However, the analysis has shown that this is not the case - for all transition sequences, the scattering rate dependence on $\Delta N$ is weak.

As an illustration, Figure 9 shows the dependence of the transition rate on the transition sequence when the initial and final Landau levels of the electron going up the ladder of Landau levels are fixed (the electron moves from level n=4 to level n=5) and the initial Landau level of the second electron of the scattering pair moves down the ladder of Landau levels. (the electron moves from level *n* to *n-1*, where *n* decreases from 4 to 1 while $\Delta N$ increasing from 0 to 3 ). In this case, according to the "concept of zeros", the decrease of *n* and the increase of $\Delta N$ works towards reducing the transition matrix element. As can be seen, the dependence on it is slightly increased compared to the situation presented in Figure 8. Nevertheless, this dependence is relatively weak.

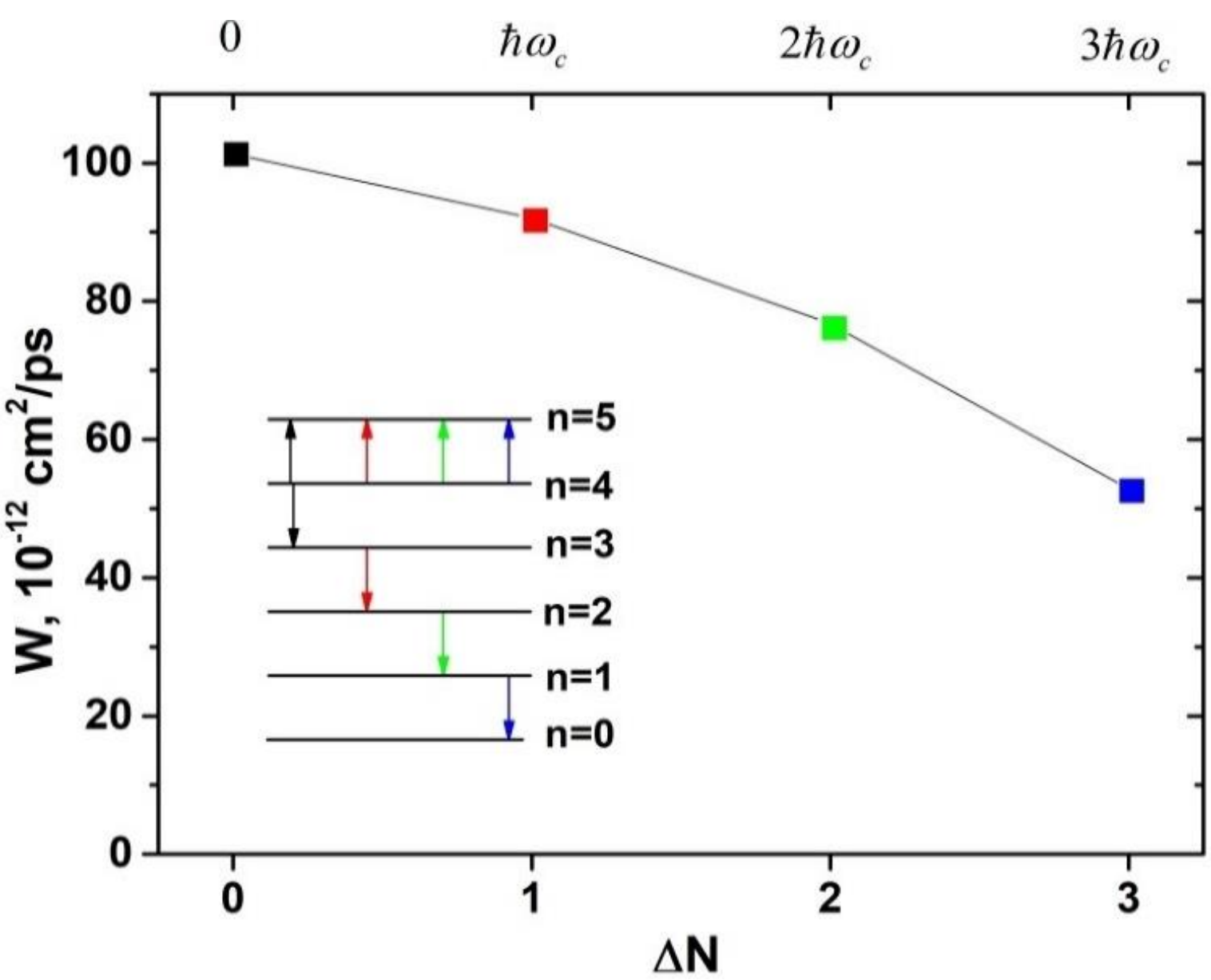

Figure. 9. Scattering rates for transitions $\{(1,4)\rightarrow(1,5)\,\&\,(1,4-\Delta N)\rightarrow(1,4-\Delta N-1)\}$ as a function of $\Delta N$.

The relatively large value of the scattering rate for $\Delta N \neq 0$ transitions is of great importance for the physical picture of relaxation in the Landau level system because of its specific mechanism. This mechanism consists of electron-electron scattering processes delivering electrons to the optical phonon level from the underlying Landau levels. Having reached Landau levels lying above or near the optical phonon energy, the electrons transfer energy to the optical vibrations of the crystal lattice in electron-phonon scattering processes [12].

The process of energy relaxation can be understood through two parallel mechanisms. The first involves the redistribution of electrons among Landau levels due to electron-electron scattering. This creates a quasi-Boltzmann distribution with an effective temperature significantly higher than that of the crystal lattice. The second mechanism is the promotion of electrons to higher Landau levels, also caused by electron-electron scattering. This is followed by the emission of optical phonons, which help transfer the electron subsystem's excitation energy to the crystal lattice.

Electron-electron scattering processes with $\Delta N \neq 0$ lead to intense transitions, as electrons at the upper levels interact with those in the highly populated lower Landau levels. This interaction significantly accelerates the thermalization of the electron subsystem [14]. Simultaneously, these transition processes "pull away" electrons from the upper Landau levels, which weakens the flow of electrons to the optical phonon level. As a result, the relaxation time strongly depends on the population of the lowest Landau level (n=0)[14].

To emphasize the significance of the $\Delta N \neq 0$ transitions in relaxation processes, we present an example in Figure 10 that illustrates the time dependence of the excitation energy of the electron subsystem. This data is calculated with and without considering the $\Delta N \neq 0$ transitions. Comparing these curves shows that neglecting the transitions leads to a substantial slowdown in energy relaxation, causing the relaxation time to increase by more than an order of magnitude, rising from 0.16 ns to 2.44 ns.

We also note that while the rate of transitions with $\Delta n > 1$ is several times lower than the rate of transitions with $\Delta n = 1$, the nonlinear nature of electron-electron scattering causes transitions with larger $\Delta n$ to influence the quantitative aspects of the relaxation kinetics significantly. This effect is illustrated in the inset of Figure 10, which shows the time evolution of the excitation energy of the system. One graph considers all $\Delta n$ transitions, while the other only includes those with $\Delta n = 1$. The results demonstrate that accounting for transitions with $\Delta n > 1$ leads to a noticeable decrease, approximately 25%, in the energy relaxation time.

We emphasize that if only transitions with the highest rates (transitions with $\Delta N = 0$ and $\Delta n = 1$) are included in the relaxation kinetics, the error in determining the relaxation time increases significantly. Precisely, the relaxation time is calculated to be 17.5 ns instead of the accurate 0.16 ns when all transitions are considered.

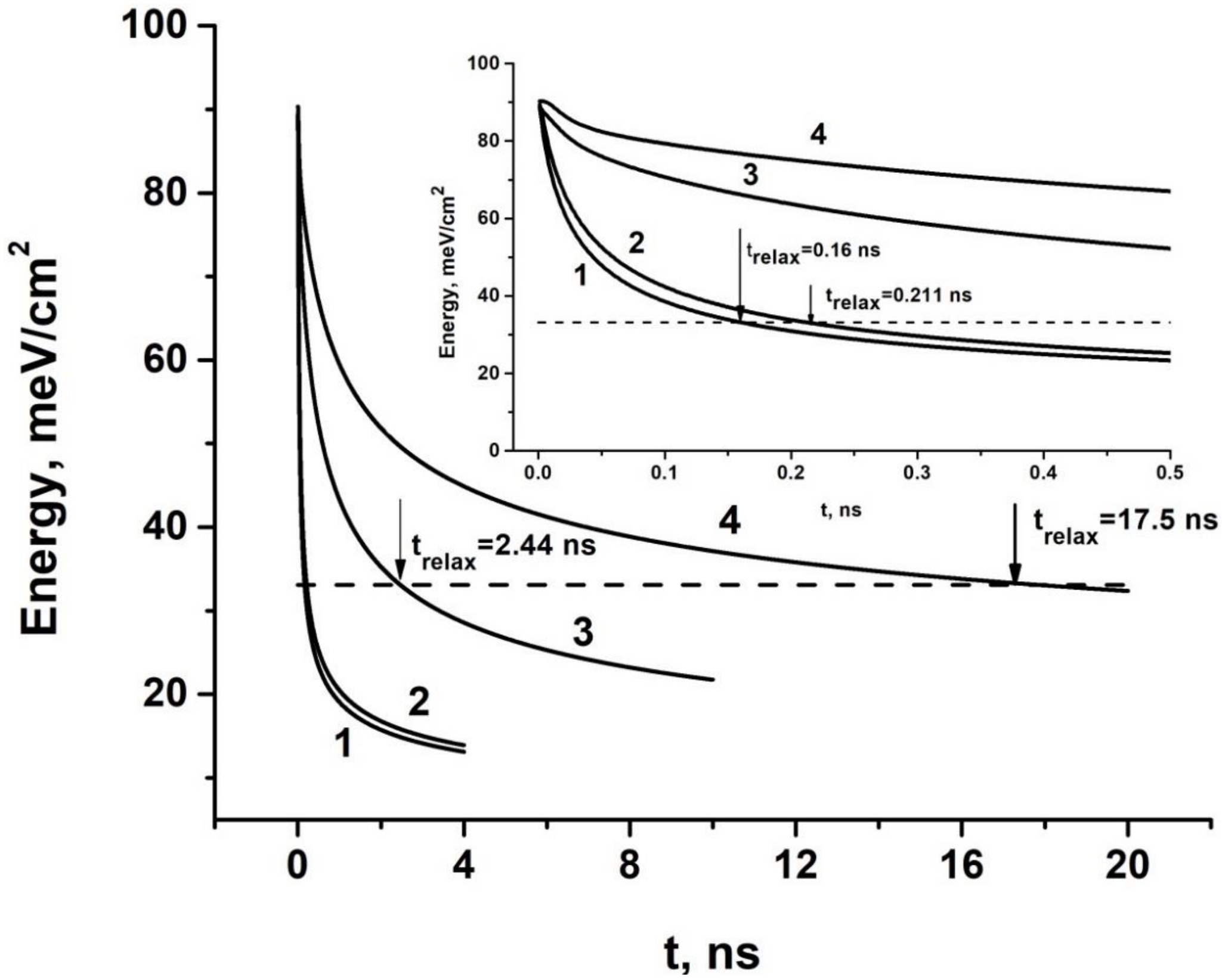

Figure 10. Dependence of the excitation energy of the electronic subsystem on time. Magnetic field B=3.5 T (at this value, the Landau level (1,6) energy is close to the optical phonon energy). Crystal lattice temperature 4.2 K. Electron density in quantum well $1.5 \cdot 10^{10}$ cm$^{-2}$. Nonequilibrium is created by the instantaneous excitation of $5 \cdot 10^{9}$ cm$^{-2}$ electrons to the Landau (1,3) level. 1 –all electron-electron scattering processes from Landau levels lying below the optical phonon energy were included; 2 – only scattering transitions with $\Delta n = 1$ were considered; 3 –only transitions with $\Delta N = 0$ were considered (but with all possible $\Delta n$); 4 – only transitions with $\Delta N = 0$ and $\Delta n = 1$ were considered. The horizontal dashed line shows the value of the excitation energy, which is e times smaller than the initial value.

We also observe that for the lowest subbands, there is a relatively small decrease in the scattering rate as the quantum well width increases (see Fig. 11). This phenomenon occurs for a reason similar to the reduction in exciton binding energy within a quantum well as the well width increases [29]. As the width of the well grows, the localization length of the wave function also increases, causing the probability density of the electron's position to become more uniformly distributed over this length. Consequently, the average distance between electrons increases, which results in a decreased scattering probability.

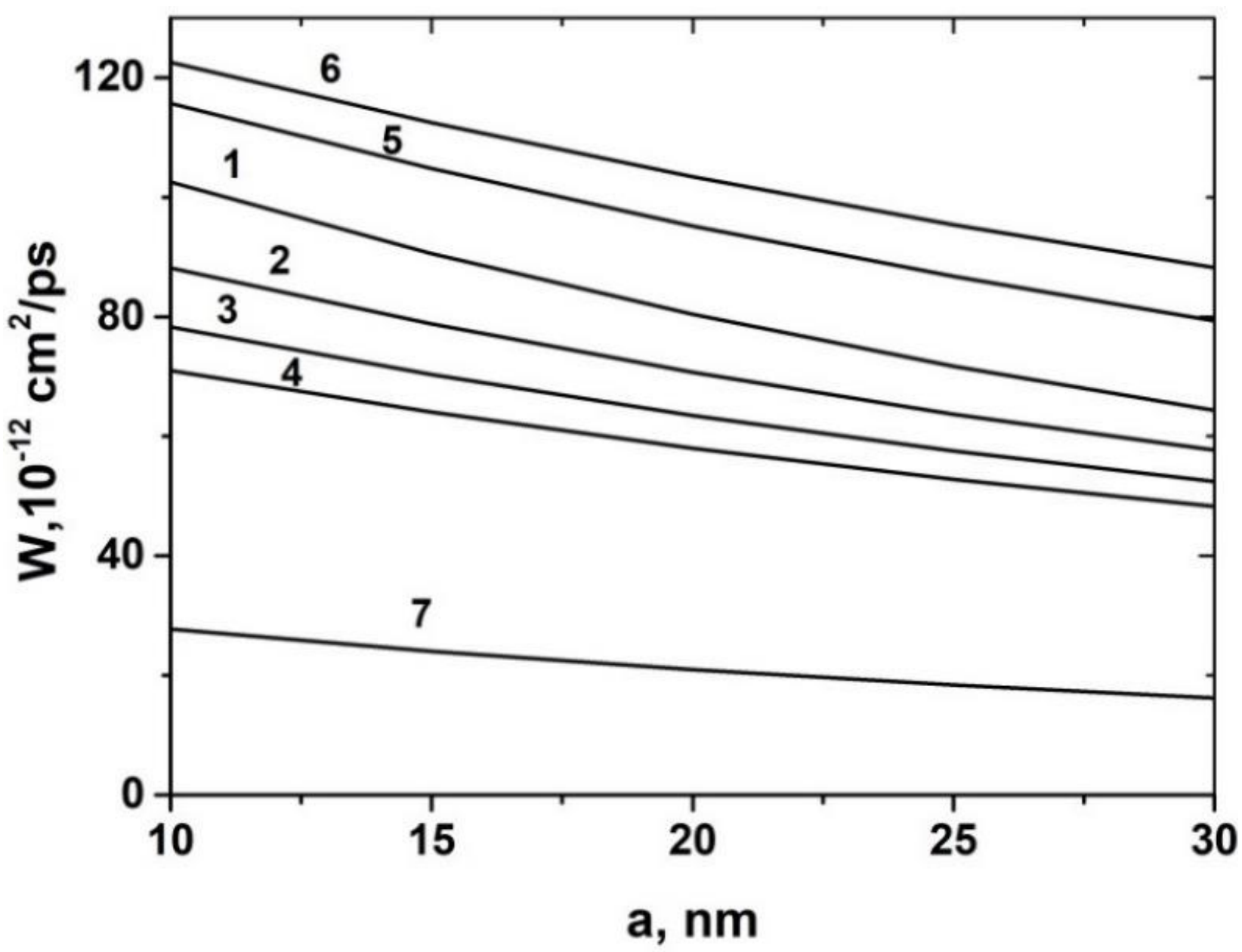

Figure 11. Dependence of the scattering rate on the quantum well width. The data are given for the quantum well GaAs/$Al_{0.3}Ga_{0.7}As$.
1 – (n=1, ΔN=0, Δn=1); 2- (n=1, ΔN=1, Δn=1); 3- (n=1, ΔN=2, Δn=1);
4- (n=1, ΔN=3, Δn=1); 5 – (n=2, ΔN=0, Δn=1); 6 – (n=3, ΔN=0, Δn=1);
7 – (n=3, ΔN=0, Δn=2).

In conclusion, we note that our calculations accounted for the nonparabolicity of the subbands. We considered band nonparabolicity according to the theory outlined in reference [35], applying second-order perturbation theory and including terms in the dispersion law expansion up to the fourth order of the wave vector. We considered transitions from Landau levels that are below the optical phonon energy up to magnetic field values where only two Landau levels of the lower subband remain below the optical phonon level. We found no significant effect of the band nonparabolicity on the electron-electron scattering rates.

## 5. Conclusion

A model of electron–electron scattering in a quantum well in a quantizing magnetic field inclined with respect to the structure layers has been developed.

Through analytical transformations, the integral multiplicity in the expression for the electron–electron scattering rate between Landau levels in the quantum well has been significantly reduced. This has allowed us to calculate the "full" electron–electron scattering rate matrix and analyze all types of transitions, both intrasubband and between Landau levels of different subbands. A rather weak dependence of the intrasubband electron–electron scattering rate on the electron energy difference in the initial state was observed, such that the scattering rates for transitions in which the electrons initially occupy the same level are close to the rates for transitions in which the electrons initially occupy levels that are quite far apart along the Landau level ladder.

It was shown that for intrasubband scattering, there is a rapid monotonic decrease in the rate with increasing energy transfer in the scattering event.